\documentclass{article}
\usepackage{spconf,amsmath,graphicx}
\usepackage{amssymb,amsfonts,amsthm}
\usepackage{euscript}
\usepackage[applemac]{inputenc}
\usepackage[T1]{fontenc}
\usepackage{ifpdf}

\usepackage[english]{babel}
\title{RNYI INFORMATION MEASURES FOR SPECTRAL CHANGE DETECTION}
\name{Marco Liuni$^1_2$, Axel R\"obel$^2$, Marco Romito$^1$, Xavier Rodet$^2$\thanks{This work is supported by grants from Region Ile-de-France.}}
\address{$^1$: Universit di Firenze, Dip. di Matematica U. Dini \\ http://www.math.unifi.it/\\
$^2$: IRCAM - CNRS STMS, Analysis/Synthesis Team \\http://anasynth.ircam.fr/}
\newcommand{\beq}{\begin{equation}}
\newcommand{\eeq}{\end{equation}}

\newtheorem{lem}{Lemma}[section]

\newtheorem{prop}[lem]{Proposition}

\begin{document}
%\ninept
%
\maketitle
\begin{abstract}
Change detection within an audio stream is an important task in several domains, such as classification and segmentation of a sound or of a music piece, as well as indexing of broadcast news or surveillance applications. In this paper we propose two novel methods for spectral change detection without any assumption about the input sound: they are both based on the evaluation of information measures applied to a time-frequency representation of the signal, and in particular to the spectrogram. The class of measures we consider, the Rnyi entropies, are obtained by extending the Shannon entropy definition: a biasing of the spectrogram coefficients is realized through the dependence of such measures on a parameter, which allows refined results compared to those obtained with standard divergences. These methods provide a low computational cost and are well-suited as a support for higher level analysis, segmentation and classification algorithms.

%The abstract should appear at the top of the left-hand column of text, about
%0.5 inch (12 mm) below the title area and no more than 3.125 inches (80 mm) in
%length.  Leave a 0.5 inch (12 mm) space between the end of the abstract and the
%beginning of the main text.  The abstract should contain about 100 to 150
%words, and should be identical to the abstract text submitted electronically
%along with the paper cover sheet.  All manuscripts must be in English, printed
%in black ink.
\end{abstract}
\begin{keywords}
Change detection, spectral entropy, Kullback divergence, Rnyi entropies, segmentation
\end{keywords}
\section{Introduction}
\label{sec:intro}

The detection of spectral changes within an audio signal can be performed according to many different criteria, depending on the applications; the key point is what kind of spectral change has to be considered significant. A typical problem in audio classification is to identify signal segments with different contents, for example when analyzing a radio stream to separate speech, music or mix of them; another type of problem is speaker change detection, which typically occurs when indexing audio recording of conferences, interviews or lectures. In either case we have to perform a segmentation and a classification, but the interesting spectral changes are completely different. The point of view we consider is at the signal level, since our research is about adaptive resolution methods for analysis, transformation and re-synthesis of a sound.

The use of information measures to evaluate the features of a time-frequency representation of a signal is frequent in the literature: Shannon entropy is applied to evaluate the concentration of the representation seen as a probability distribution, and the derived divergence measures \cite{Lin02} are employed to identify variations within the representation. 

%The discrete spectrogram of a signal is a sampling of the square of its continuous version $\mathrm{PS}_f = |\mathrm{S}f|^2 ~,$
%where $f$ is a signal, and $\mathrm{S}f$ is the STFT of $f$ through a window function. Such a sampling is obtained according to a regular lattice: given a time step $a$ and a frequency step $b$ we write $\{u_n\}_{n\in\m{Z}} = an$ and $\{\xi_k\}_{k\in\m{Z}} = bk$; these two sequences generate the nodes of the time-frequency lattice $\Lambda$ where the spectrogram values are sampled,
%\beq\label{disc_spec_def}  \mathrm{PS}_f[n,k] = |\mathrm{S}f[u_n,\xi_k]|^2~.\eeq 

The representation we consider is the spectrogram of the signal: through a normalization which gives a unitary sum, we consider the discrete spectrogram in a finite time interval as a probability distribution, and we can apply typical information measures to evaluate its concentration in the time-frequency plane. Fixing the signal $f$, we write $\mathrm{PS}_{m} = \{\mathrm{PS}_f[m,k],~k=1,...,N\}$ to indicate the $m$-th analysis frame in the discrete spectrogram $\mathrm{PS}_f$ of $f$, where the \emph{FFT size} $N$ is the finite number of sample frequencies considered. Given two normalized analysis frames $\mathrm{PS}_{1}$ and $\mathrm{PS}_{2}$, the \emph{Kullback} $K$ divergence  \cite{Lin02} is usually employed to have a measure of their difference: a spectral change is detected whenever $K(\mathrm{PS}_{1},\mathrm{PS}_{2})$ is larger than a chosen threshold. A refinement of this method (see for example \cite{Ba03}) provides a better robustness to false alarms defining a \emph{mean spectrum}  $\mathrm{PS}_{mean}$ and comparing its divergence with the new analysis frame.\\ 

%. The \emph{Kullback} $I$ and $J$ divergence measures are derived from the Shannon entropy \cite{Re61} as follows, where we assume $0 \log 0 = 0$ and $\log\frac{0}{0} = 0~.$ The $I$ \emph{directed divergence} is
%
%\beq\label{I_div_def} I(\mathrm{PS}_{1},\mathrm{PS}_{2}) = \sum_{k = 1}^N \mathrm{PS}_{1}[k] \log\frac{\mathrm{PS}_{1}[k]}{\mathrm{PS}_{2}[k]}~,\eeq 
%which is nonnegative and additive but not symmetric; a symmetric extension of this measure is given by 
%
%\beq\label{J_div_def} J(\mathrm{PS}_{1},\mathrm{PS}_{2}) = \sum_{k = 1}^N (\mathrm{PS}_{1}[k] - \mathrm{PS}_{2}[k]) \log\frac{\mathrm{PS}_{1}[k]}{\mathrm{PS}_{2}[k]}~,\eeq 
%where $\mathrm{PS}_{1}$ has to be absolutely continuous with respect to $\mathrm{PS}_{2}$ for $I$ to be defined while $\mathrm{PS}_{1}$ and $\mathrm{PS}_{2}$ have to be equivalent in the definition of $J$. The \emph{$K$ directed divergence} \cite{Lin02} is an alternative well suited for difference measures; it is defined as
%
%\beq\label{K_div_def} K(\mathrm{PS}_{1},\mathrm{PS}_{2}) = \sum_{k = 1}^N \mathrm{PS}_{1}[k] \log\frac{\mathrm{PS}_{1}[k]}{\frac{1}{2}\mathrm{PS}_{1}[k] + \frac{1}{2}\mathrm{PS}_{2}[k]}~,\eeq 
%so we have $K(\mathrm{PS}_{1},\mathrm{PS}_{2})\geq 0$ and the equality is attained only if $\mathrm{PS}_{1} = \mathrm{PS}_{2}$. The last two measures are both derived from the $I$ one, as $ J(p,q) = I(p,q) + I(q,p)$ and $K(p,q) = I(p,\frac{1}{2}p + \frac{1}{2}q)$.

%If we consider for example the last measure introduced, 

The first method we propose is a straight extension of the one just described: we consider the divergence measure derived from the \emph{Rnyi entropy} \cite{Re61} instead of the $K$ directed divergence, allowing a tuning of the detection criteria thanks to the dependance of the measure on a parameter. The second method is not based on divergence but on Rnyi entropy itself, exploiting one of its fundamental property: the entropy of a union of probability distributions can be evaluated considering the entropy values of the individual distributions. Since we do consider analysis frames as probability distributions, this property can be used to establish the expected entropy value of a certain signal segment when the following frame is added: if the actual value differs significantly from the expected one, the last frame is considered to contain a spectral change.\\

This kind of algorithm does not need acoustic models to refer to, nor data training: a certain metric is evaluated in a given space \cite{KS00}. The information measures we take into account can be applied on several different representation of the signal: in \cite{SJ97} the $K$ divergence is used in a GMM framework instead of on the spectrogram. In several approach, for example in \cite{Fo02}, difference measures are calculated as a first step which gives a suitable analysis for segmentation and classification purposes: for all these algorithms, the class of measures we introduce could ameliorate the detection performances as they allow a further parameter of choice, while still including the $K$ divergence for a given value of the parameter.\\

In the next section we give the essential properties and definitions of the measures considered, then we describe the biasing obtained with the parameter introduced. Finally we present our algorithms and give some examples: we use a speech fragment to compare the detection with the one given by the $K$ divergence measure; we take as a reference the segmentation given on the same signal by an HMM-based phoneme segmentation method \cite{La08}, and the voiced-unvoiced classification obtained with a PSOLA-based algorithm \cite{MVV06}. Our results are interesting as the methods provide a refined adjustable detection, despite of their substantial plainness and low computational cost.

%These guidelines include complete descriptions of the fonts, spacing, and
%related information for producing your proceedings manuscripts. Please follow
%them and if you have any questions, direct them to Conference Management
%Services, Inc.: Phone +1-979-846-6800 or Fax +1-832-426-7760 or email
%to icassp2011@securecms.com.

\section{RNYI ENTROPIES AND INFORMATION MEASURES}
\label{sec:rnyi}

Given a finite probability density $p$ and a rational number $\alpha \geq 0$, the Rnyi entropy of $p$ is defined as follows,

\beq\label{rent_disc_def} \mathrm{H}_{\alpha}[p] = \frac{1}{1 - \alpha} \log_2 \sum_{k = 1}^N p^{\alpha}[k]~,\eeq
where $p$ is in square brackets as we are considering the measure on discrete densities; as $\alpha$ tends to one this measure converges to the Shannon entropy, which is therefore included in this larger class. General properties of Rnyi entropies can be found in \cite{Re61}, \cite{BS93} and \cite{Zy04}; in particular, $\mathrm{H}_{\alpha}(P)$ is a non increasing function of $\alpha$, so
$\alpha_1 < \alpha_2 \Rightarrow \mathrm{H}_{\alpha_1}(P)\geq \mathrm{H}_{\alpha_2}(P)~.$ Moreover, for every order $\alpha$ the Rnyi entropy $\mathrm{H}_{\alpha}$ is maximum when $P$ is uniformly distributed, while it is minimum and equal to zero when $P$ has a single non-zero value. As we are working with finite discrete densities we can also consider the case $\alpha = 0$ which is simply the logarithm of the number of elements in $p$; as a consequence $\mathrm{H}_0[p] \geq \mathrm{H}_{\alpha}[p]$ for every admissible order $\alpha$. Given a second finite probability density $q$ of the same length, if $p$ and $q$ have exactly the same zeros the \emph{Rnyi information} \cite{Re61} is defined as follows,

\beq\label{ren_div_def} I_{\alpha}(q,p) = \frac{1}{\alpha - 1} \log_2 \sum_{k = 1}^N \frac{q^{\alpha}[k]}{p^{\alpha - 1}[k]}~,\eeq 
and it tends to the Kullback $I$ divergence \cite{Lin02} as $\alpha$ tends to one. We can thus consider this class of measures to obtain different divergences as for the Kullback $I$ one, and apply them to the spectrogram frames: as long as we can give an interpretation to the $\alpha$ parameter, this class of measures offers a largely more detailed information about the time-frequency representation of the signal.

\subsection{Biasing spectral coefficients through the $\mathbb{\alpha}$ parameter}
\label{ssec:sub_alpha}
\begin{figure}[htb]

\begin{minipage}[b]{1.0\linewidth}
  \centering
  \centerline{\includegraphics[width=9.5cm]{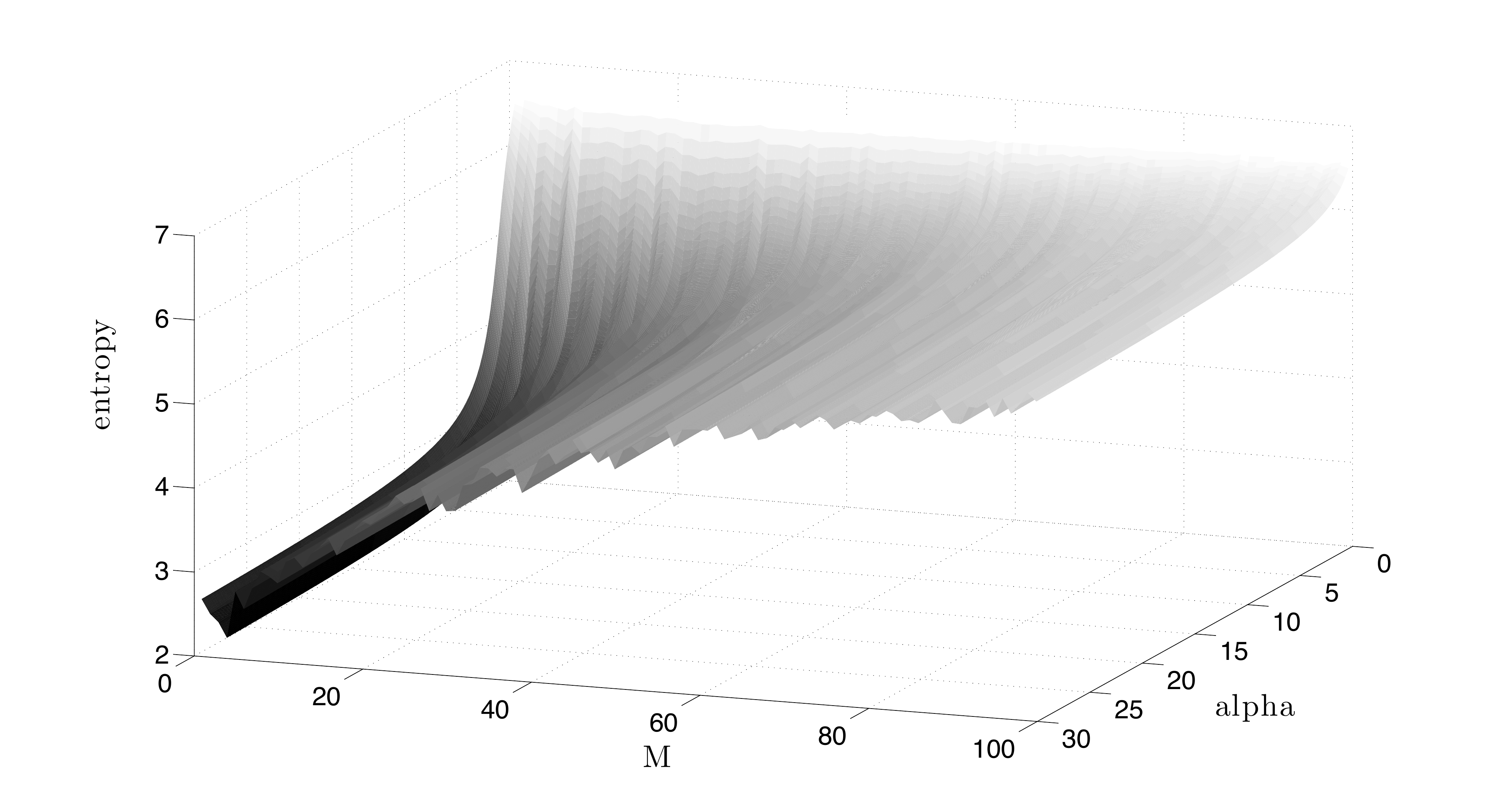}}\medskip
\vspace{-0.5cm}
\end{minipage}

\caption{Rnyi entropy evaluations of the $U_M$ vectors with varying $\alpha$.}
\label{fig:test_alpha}
\end{figure}

To show the biasing introduced on the spectral coefficients by the $\alpha$ parameter we consider a simplified model of a spectrogram composed by a variable amount of large and small coefficients. We realize a vector $U$ of length $N = 100$ generating numbers between 0 and 1 with a normal random distribution; then we consider the vectors $U_M,~1\leq M\leq N$ such that

\begin{displaymath} U_M[k] = \left\{ \begin{array}{ll}
U[k] & \textrm{if $k\leq M$}\\ \frac{U[k]}{20} & \textrm{if $k > M$}
\end{array} \right. \end{displaymath}
and then normalize to obtain a unitary sum. We then apply Rnyi entropy measures with $\alpha$ varying between 0 and 30: as we see from figure \ref{fig:test_alpha}, there is a relation between $M$ and the slope of the entropy curves for the different values of $\alpha$.

For $\alpha = 0$, $\mathrm{H}_0[U_M]$ is the logarithm of the number of non-zero coefficients and it is therefore constant; when $\alpha$ increases, we see that densities with a small amount of large coefficients gradually decrease their entropy. This means that increasing $\alpha$ we emphasize the difference between the entropy values of a peaky distribution and that of a nearly flat one. In the next section we will give an example of the exploiting of this important property, but care should be taken when applying this criterium: small coefficients in a spectrogram include signal components of weak amplitude as well as noise; choosing an extremely small $\alpha$ the change detection robustness to noise level significantly decreases.

\subsection{The entropy prediction method}
\label{ssec:ent_pred}
The second method we introduce is not based on a divergence criterium, but on entropy itself. We first give the definition of Rnyi entropy for the case of distribution obtained with a discretization of their continuous version \cite{BF01}: let $\mathrm{PS}_f$ be a normalization with unitary sum of a discrete spectrogram, then the Rnyi entropy of $\mathrm{PS}_f$ is
\beq\label{ren_ent_disc} \mathrm{H}_{\alpha}[\mathrm{PS}_f ] =  \frac{1}{1-\alpha}\log_2 \sum_{n,k}(\mathrm{PS}_f[n,k])^{\alpha}  + \log_2(ab)~,\eeq
where $k$ varies between 1 and the FFT size $N$ while $n$ varies in the time interval where the evaluation has to be performed, according to the time grid. The term $ \log_2(ab)$ takes into account the time and frequency steps $a$ and $b$ of the lattice $\Lambda$ used to sample the continuous spectrogram: this guarantees the stability of the discrete entropy when changing the hop and the FFT sizes, as long as the sampling grid is dense enough in the time-frequency plane. For the entropy of a single analysis frame we write $\mathrm{H}_{\alpha}[\mathrm{PS}_f ] =  \mathrm{H}_{\alpha}[\mathrm{PS}_m]$ as above, where $m$ is the time index of the analysis frame considered; for $L$ different analyses frames, we write $\mathrm{H}_{\alpha}[\mathrm{PS}_f ] = \mathrm{H}_{\alpha}[\mathrm{PS}_{m},...,\mathrm{PS}_{m + L}]$ to focus on the individual vectors. The following properties are straightforward by the definitions.

\begin{prop}[Rnyi entropy prediction]
Consider a spectrogram $\mathrm{PS}_f$ and a rational number $\alpha \geq 0$.
\begin{enumerate}
\item[(i)] Let $\mathrm{PS}_{m}$ be an analysis frame in $\mathrm{PS}_f$; if $\mathrm{PS}_{k}$ is obtained rearranging the elements of $\mathrm{PS}_{m}$, then 
\beq\label{iso_ent_1} \mathrm{H}_{\alpha}[\mathrm{PS}_{m}] = \mathrm{H}_{\alpha}[\mathrm{PS}_{k}] = \mathrm{H}~,\eeq
\beq\label{iso_ent_2} \mathrm{H}_{\alpha}[\mathrm{PS}_{m},\mathrm{PS}_{k}] = \mathrm{H} + 1~.\eeq
\item[(ii)]In general, if $\mathrm{PS}_{m + 1},...,\mathrm{PS}_{m + L}$ are obtained rearranging the elements of $\mathrm{PS}_{m}$, than
\beq\label{iso_ent_3} \mathrm{H}_{\alpha}[\mathrm{PS}_{m},...,\mathrm{PS}_{m + L}] = \mathrm{H} + \log_2(L + 1)~.\eeq
\end{enumerate}
\end{prop}
As a rearrangement we mean a reordering of the frame coefficients, thus including the case of equality between frames. The idea of our method is that given the entropy of a certain signal segment $\mathrm{H}_{\alpha}(\mathrm{PS}_m,...,\mathrm{PS}_{m + L})$ composed by $L$ contiguous frames, we can predict $\mathrm{H}_{\alpha}(\mathrm{PS}_m,...,\mathrm{PS}_{m + L +1})$ supposing the new frame to be spectrally coherent and thus iso-entropic with the previous ones. If on the other hand the entropy value of the new segment largely differs from the predicted value, we assume the new frame to be incoherent with the previous and so a spectral change is detected. There is here a strong assumption concerning the equivalence between the concept of spectral coherence and the fact that two frames are obtained with a rearrangement of their elements; according to the specific needs in the applications, the detection criteria can be based on variations of the property \eqref{iso_ent_3} to take into account different definitions of spectral coherence: for example, considering a set of admissible operations on the analysis coefficients in relation with the entropy variation that they provide.

\section{ALGORITHMS AND EXAMPLES}
\label{sec:algo_ex}

We show here an application of the detection algorithms with the measures defined: the first algorithm we analyze has the same operations for the $K$ divergence and Rnyi information \eqref{ren_div_def}: we calculate the spectrogram of a signal with a 1024-samples Hamming window, 768-samples overlap and 2048-points FFT size; we obtain a mean spectrum taking the first 20 analysis frames, and calculate the divergence of the next frame with respect to the mean spectrum. Once we have the first divergence value, we shift the mean spectrum of one analysis frame and consider the following 20 frames, then calculate the divergence between the new mean spectrum and the following frame. At this point, if the ratio between the last divergence value and the previous exceeds a certain threshold, a change is detected at the incoming frame; otherwise the procedure goes on.
The second algorithm is a variation of the first one based on entropy prediction: once obtained the spectrogram of the signal, we calculate the Rnyi entropy of the vector composed of its first 6 analysis frames; then we consider the next frame and set the predicted entropy value according to \eqref{iso_ent_3}. We calculate the actual entropy of the vector obtained adding the new frame to the previous ones, and if the ratio between this value and the predicted one exceeds a certain threshold, a change is detected. Then the procedure goes on as in the previous case.\\

\begin{figure*}[ht]
\begin{minipage}[b]{1.0\linewidth}
  \centering
  \centerline{\includegraphics[width=10cm, bb = 3 67 607 714]{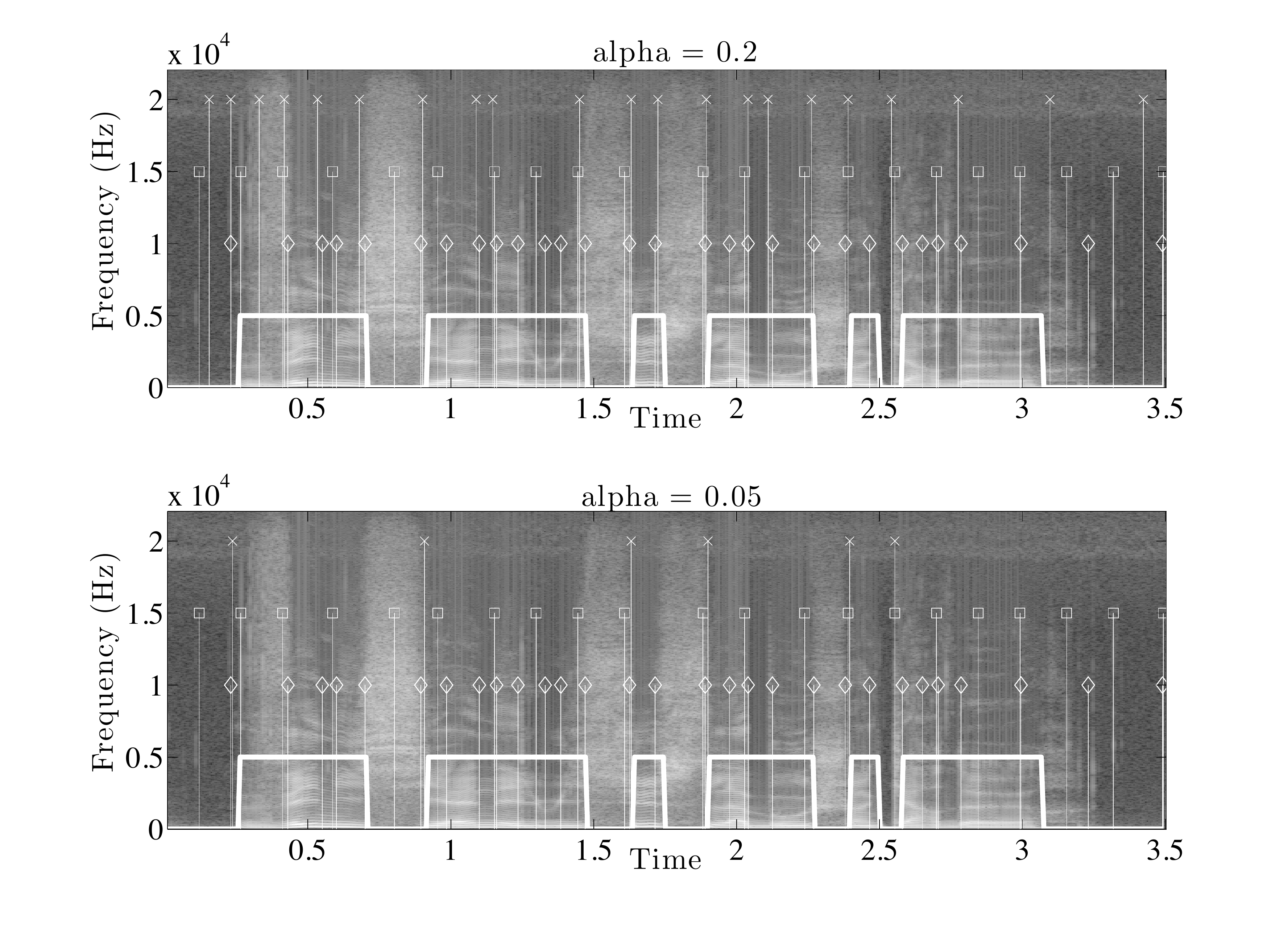}}\medskip %3 257 607 534
\vspace{-0.5cm}
\end{minipage}
\caption{\label{fig:voix_x}{\it Detections obtained with different methods on a speech fragment in French language; \textbf{cross markers}: Rnyi entropy prediction method, on top with $\alpha = 2$, at the bottom with $\alpha = 1.1$; \textbf{square markers}: $K$ divergence; \textbf{diamond markers}: HMM-based phoneme segmentation method; \textbf{bold line}: PSOLA voiced-unvoiced classification, 0 is unvoiced. }}
\end{figure*}

The Rnyi prediction shows a slightly better accuracy at the price of a higher computational cost; this is due to the larger dimensions of the vectors managed in the entropy calculus. The tuning of the $\alpha$ parameter gives interesting results: as seen in figure \ref{fig:test_alpha}, higher values rise the difference between the entropies of a peaky distribution and a flat one; thus we expect in general a more refined detection increasing $\alpha$, leaving the threshold unchanged. The signal we analyze is a speech fragment of a mail voice in French language, \emph{Vnitienne et lui suce la bouche un quart d'heure}. We assume two references: an automatic phoneme segmentation for French language based on Hidden Markow Model \cite{La08}, and a voiced-unvoiced classification obtained with a PSOLA-based algorithm \cite{MVV06}: they identify the major spectral changes in this kind of signal, so we expect our detection to confirm them. We are not interested in whether a marker belongs to one selection or the other, as this could be established in a later classification step. As we see at the top of figure \ref{fig:voix_x}, the Rnyi prediction with $\alpha = 0.2$ identifies all the voiced-unvoiced transitions in both senses except at time 2.5, and a large part of phonemes. If we need a less refined detection, setting the $\alpha$ parameter to 0.05 (bottom of figure \ref{fig:voix_x}) preserves the detection of all the unvoiced-voiced transitions, while discarding all the phonemes and the voiced-unvoiced transitions. Both the measures provide a better detection with respect to the $K$ divergence, which shows a higher number of unexpected markers.

\vfill
\pagebreak

%\section{FOOTNOTES}
%\label{sec:foot}
%
%Use footnotes sparingly (or not at all!) and place them at the bottom of the
%column on the page on which they are referenced. Use Times 9-point type,
%single-spaced. To help your readers, avoid using footnotes altogether and
%include necessary peripheral observations in the text (within parentheses, if
%you prefer, as in this sentence).

%\section{COPYRIGHT FORMS}
%\label{sec:copyright}
%
%You must include your fully completed, signed IEEE copyright release form when
%you submit your paper. We {\bf must} have this form before your paper can be
%published in the proceedings.  The copyright form is available as a Word file,
%a PDF file, and an HTML file. You can also use the form sent with your author
%kit.

%\section{REFERENCES}
%\label{sec:ref}
%
%List and number all bibliographical references at the end of the paper.  The references can be numbered in alphabetic order or in order of appearance in the document.  When referring to them in the text, type the corresponding reference number in square brackets as shown at the end of this sentence \cite{C2}.

% References should be produced using the bibtex program from suitable
% BiBTeX files (here: strings, refs, manuals). The IEEEbib.bst bibliography
% style file from IEEE produces unsorted bibliography list.
% -------------------------------------------------------------------------
\bibliographystyle{IEEEbib}
\bibliography{general_bib}

\end{document}